\documentclass{article}
\usepackage{amssymb}
\usepackage{amsfonts}
\usepackage{amsmath}

\input{tcilatex}
\begin{document}

\title{\textbf{Scalar} \textbf{Casimir effect between two concentric spheres}%
}
\author{Mustafa \"{O}zcan\thanks{%
Electronic address: ozcanm@trakya.edu.tr and ozcanm99@gmail.com} \\
Department of Physics, Trakya University 22030 Edirne, Turkey}
\maketitle

\begin{abstract}
The Casimir effect giving rise to an attractive force between the closely
spaced two concentric spheres that confine the massless scalar field is
calculated by using a direct mode summation with contour integration in the
complex plane of eigenfrequencies. We devoleped a new approach appropriate
for the calculation of the Casimir energy for spherical boundary conditions.
The Casimir energy for a massless scalar field between the closely spaced
two concentric spheres coincides with the Casimir energy of the parallel
plates for a massless scalar field in the limit when the dimensionless
parameter $\eta ,$ $(\eta =\frac{a-b}{\sqrt{ab}}$ where $a$ $(b)$ is inner
(outer) radius of sphere), goes to zero. The efficiency of new approach is
demonstrated by calculation of the Casimir energy for a massless scalar
field between the closely spaced two concentric half spheres.

PACS number(s): 03.70.+k, 12.20.DS, 11.10.Gh
\end{abstract}

\section{\protect\bigskip INTRODUCTION}

\bigskip \qquad In 1948 Casimir \cite{Cas} had shown that two uncharged
perfectly conducting parallel plates attracted each other. The main idea
under this attractive force is that changes in infinite vacuum energy of the
quantized electromagnetic field can be finite and observable \cite{Spaar,
Lam,mohideen}. The sign of Casimir energy depends on manifolds of different
topology, geometry, the shapes and compositions of objects as well as the
boundary and curvature \cite{Mozcan2, Mozcan3, Miltonbook}. The Casimir
effect has been studied in different areas of theoretical physics and an
important results may be found in detalied reviews \cite{Gunter, Mosbook,
Mos1}.

It is well-known fact that the negative sign of the Casimir energy between
the plates for the electromagnetic field produces an attractive force.
Stimulated by this event, Casimir hoped that one could construct an electron
model as a perfectly conducting charged spherical shell. However, Boyer \cite%
{Boyer} showed that the quantum electromagnetic vacuum energy of a
conducting spherical shell is positive which means that the force on the
shell is outward and depends only on the radius of the spherical shell. This
outward force was opposite to what Casimir had expected. Later, Milton et.
al. \cite{Milton1} confirmed Boyer result by using Green's functions. More
recently, the direct mode summation technique and the Casimir energy
calculations for the spherical shell has been advanced by \cite{Nest1, Hag}.
Nowadays, the Casimir effect between two objects has also been considered 
\cite{bordag,emig1,emig2,emig3,kenneth,zaheer,rahi}. Several different
methods have been developed to calculate the Casimir effect beyond the
proximity force approximation. One of the pioneering study is the work by M.
Bordag \cite{bordag}, where he developed to compute the small separation
leading order terms of the Casimir interaction between a cylinder and a
plane by using the path integral approach. Moreover, the exact result for
the Casimir interactions between a finite number of compact objects of
arbitrary shape and separation have been developed by using the functional
determinant or multiple scattering approach \cite%
{emig1,emig2,emig3,kenneth,zaheer,rahi}. Besides, the Casimir force problems
of two concentric spheres \cite{Brevik1, Miltao, teo1, IHD, Mozcan1} and two
concentric cylinders \cite{Jiang, Tatur, Mazzitelli} have also been
calculated by using different regularization method recently. \ 

In this work we calculate the Casimir energy for the massless scalar field
in a nontrivial two smoothly curved objects at close separation. The system
we consider is made of two compact closely spaced spherical surfaces and the
vacuum gap consists of between two concentric spheres with radii $a$ and $b$ 
$\left( b>a\right) .$ There are two motivations for us to perform the
Casimir energy calculation for a massless scalar field between the closely
spaced two concentric spheres, and between the closely spaced two concentric
half spheres \cite{Mozcan1, Funaro}. First it is interesting to find what
similarities and difference of the Casimir energy of the toplogically
similar geometry and comparing the parallel plates and the spherical shell,
respectively. The sign of the Casimir energy will decide whether the Casimir
force will be attractive or repulsive. We hope that there will be an
application of our result to special systems in nanotechnologies and
nanoelectromechanical devices. Second from the mathematical point of view
the direct mode summation approach to the our geometry has been developed.
Our aim is to show the simplicity and efficiency of the direct mode
summation by contour integration when calculating the Casimir energy for a
difficult boundary as the closely spaced two concentric spheres. The direct
evaluation of the infinite sum over all the vacuum energy eigenvalues of the
massless scalar field modes implies that the translation matrices in
multiple scattering approach is not needed in order for the finite values
for the vacuum energy to be obtained for given spherical boundary
conditions. Moreover, the improved value for the computation of the Casimir
energy for a massless scalar field between the closely spaced two concentric
half spheres we reconsider here is developed by using the Abel-Plana sum
formula for evenly spaced frequency spectrum for large argument.

The organization of the paper is as follows. In Sec. 2. The Casimir energy
of a massless scalar field subjected to spherical boundary conditions on
between the closely spaced two concentric spheres is calculated without any
approaximation techniques. This approach will be employed for the Casimir
energy between the closely spaced two concentric half spheres in Sec. 3.
Concluding remarks and discussion of the Casimir energy for a massless
scalar field in an annular region of our geometry is presented in section 4.

The units are such that $\hbar =c=1.$

\bigskip

\section{CASIMIR ENERGY BETWEEN TWO CONCENTRIC SPHERES}

\qquad We start with the spacetime metric

\begin{equation}
ds^{2}=dt^{2}-\left( dr^{2}+r^{2}d\theta ^{2}+r^{2}\sin ^{2}\theta d\phi
^{2}\right)
\end{equation}%
in spherical coordinates. Where $\theta \in \left[ 0,\pi \right] $ and $\phi
\in \left[ 0,2\pi \right] $. The wave equation that the massless scalar
field satisfies in these coordinates is given by

\begin{equation}
\square \Psi \left( t,r,\theta ,\phi \right) =0.
\end{equation}%
Where $\square $ is the D'Alembertian operator associated with the metric
given by the line element Eq. (1). Solution of equation (2) could be easily
found by using the method of separation of variables and is given as

\begin{equation}
\Psi _{\omega \ell m}\left( t,r,\theta ,\phi \right) =\sum_{\ell
,m}\,e^{-i\omega t}\,r^{-\frac{1}{2}}\;\left[ A_{\ell }\,J_{\ell
+1/2}(\omega r)\,\,+B_{\ell }\,N_{\ell +1/2}(\omega r)\right] \,Y_{\ell
m}(\theta ,\phi )\,.
\end{equation}%
Where the $Y_{\ell m}(\theta ,\phi )$ are the Spherical harmonics, and $%
J_{\ell +1/2}(\omega r)\,$\ and $\,N_{\ell +1/2}(\omega r)$ are the Bessel
functions of the first and second kind. The coefficents $A_{\ell }$ and $%
B_{\ell }$ can be determined from the boundary conditions, respectively. And 
$\ell =0,1,2,3,...\;\;$and $m=-\ell ,-\ell +1,....0,1,2,...,\ell -1,\ell $.

We now impose the boundary conditions for our geometry ($a$ ($b$)is inner
(outer) radius of sphere) that is

\begin{equation}
\Psi _{\omega \ell m}\left( t,r=a,\theta ,\phi \right) =0,\;\text{and
similarly }\Psi _{\omega \ell m}\left( t,r=b,\theta ,\phi \right) =0\;.
\end{equation}%
The eigenfunction that satisfy the boundary conditions is%
\begin{equation}
\Psi _{\omega \ell m}\left( t,r,\theta ,\phi \right) =c_{0}\sum_{\omega
,\ell ,m}\,e^{-i\omega t}\,r^{-\frac{1}{2}}\left[ J_{\ell +1/2}(\omega r)\,-%
\frac{J_{\ell +1/2}(\omega a)}{\,\,N_{\ell +1/2}(\omega a)}\,N_{\ell
+1/2}(\omega r)\right] \,Y_{\ell m}(\theta ,\phi )\,.
\end{equation}%
Where $c_{0}$ is the normalization constant and $\omega $ is the root of the
following transcendental equation

\begin{equation}
J_{\ell +1/2}(\omega b)\,\,N_{\ell +1/2}(\omega a)-J_{\ell +1/2}(\omega
a)\,\,N_{\ell +1/2}(\omega b)=0\;.
\end{equation}

\bigskip

The Casimir energy for the massless scalar field between two concentric
spheres is

\begin{eqnarray}
E_{C} &=&\frac{1}{2}\sum_{\ell =0}^{\infty }\sum_{m=-\ell }^{\ell
}\sum_{n=1}^{\infty }\;\omega _{n\ell }\;,  \notag \\
&&  \notag \\
&=&\sum_{\ell =0}^{\infty }\nu \sum_{n=1}^{\infty }\;\omega _{n\ell }~\ \ \
\ \ \ \ \ \ \ \ \text{where\ }\nu =\ell +\frac{1}{2}.
\end{eqnarray}%
Where $\omega _{n\ell }$ are eigenfrequencies which are determined by
solving the frequency equation given in Eq. (6). We need to describe the
eigenfrequencies spectrum for the Casimir energy in a nontrivial smoothly
curved objects at close separation. We note that the frequency equation
involves an infinite series since the complete solution of Bessel's equation
has series of ascending powers of $\omega $. Bessel's series equation are
convergent for all values of argument. But, when $\left\vert \omega
\right\vert $ is large, the series converge slowly. Although the series has
the converge slowly, the initial terms of such a series gives no information
about the sum. Thus Bessel functions are needed to describe the transition
for large argument. To overcome this difficulty, we need the rapidly
convergent evaluation of the Bessel's function formula. The meaning of
rapidly convergent is that the series rapidly approaches a constant, taking
the limit as $\ell \rightarrow \infty $. To this aim, we will use the
uniform asymptotic expansions of the Bessel's functions. The uniform
asymptotic expansions are useful in describing the transition of behaviour.
Moreover, after taking the limit $\omega \rightarrow \infty $ at fixed $\ell 
$ the spectrum will consist of a discrete set embedded in a continum part.
Hence, we should examine the behavior of the eigenfrequency spectrum for
large arguments at fixed $\ell .$ Thus, to carry out the summation with
respect to $\ell $ in $E_{C}$, $\ $the sum $\sum_{n=1}^{\infty }\;\omega
_{n\ell }$ \ given in equation (7) replaced by $\sum_{n=1}^{\infty }\;\omega
_{n\ell }+\sum_{n=1}^{\infty }\;\widetilde{\omega }_{n\ell }$ where $%
\widetilde{\omega }_{n\ell }$ \ is the eigenvalue spectrum of the limit $%
\omega \rightarrow \infty $ at fixed $\ell ~$\cite{Boyer}. Then, Casimir
energy which is defined by the eigenfrequency spectrum for large arguments
at fixed $\ell $ and large order as $\ell \rightarrow \infty $ can be
written as

\begin{equation}
E_{C}=\sum_{\ell =0}^{\infty }\nu \sum_{n=1}^{\infty }\;\omega _{n\ell
}+\sum_{\ell =0}^{\infty }\nu \sum_{n=1}^{\infty }\widetilde{\omega }_{n\ell
}\ \ \;.
\end{equation}

Now, to calculate the eigenfrquencies for large arguments at fixed $\nu $,
we use Hankel's asymptotic expansion \cite{Abramowitz} when $\nu $ is fixed, 
$\omega a\gg 1$ and $\omega b\gg 1$, we get

\begin{equation}
J_{\nu }\left( \widetilde{\omega }a\right) \simeq \sqrt{\frac{2}{\pi 
\widetilde{\omega }a}}\left[ \cos \left( \widetilde{\omega }a-\frac{\nu }{2}%
\pi -\frac{\pi }{4}\right) -\frac{\left( 4\nu ^{2}-1\right) }{8\widetilde{%
\omega }a}\sin \left( \widetilde{\omega }a-\frac{\nu }{2}\pi -\frac{\pi }{4}%
\right) \right]
\end{equation}

\bigskip 
\begin{equation}
N_{\nu }\left( \widetilde{\omega }a\right) \simeq \sqrt{\frac{2}{\pi 
\widetilde{\omega }a}}\left[ \cos \left( \widetilde{\omega }a-\frac{\nu }{2}%
\pi -\frac{\pi }{4}\right) +\frac{\left( 4\nu ^{2}-1\right) }{8\widetilde{%
\omega }a}\sin \left( \widetilde{\omega }a-\frac{\nu }{2}\pi -\frac{\pi }{4}%
\right) \right]
\end{equation}%
and similar expressions for $J_{\nu }\left( \widetilde{\omega }b\right) $
and $N_{\nu }\left( \widetilde{\omega }b\right) $ with $a$ interchanged for $%
b$. Putting (9) and (10) in the frequency equation given by (6), we obtain
the zeros of frequency equation are almost evenly spaced

\begin{equation}
\widetilde{\omega }_{n\ell }^{2}\simeq \left( \frac{n\pi }{b-a}\right) ^{2}+%
\frac{\nu ^{2}}{ab}\ \ \ \text{where}\ \ \ n=1,2,3,4,5,6,...\ \ .
\end{equation}

\bigskip

The frequencies equation $\omega _{n\ell }$ the first sum is given in Eq.
(8) as the uniform asymptotic expansions of the Bessel functions at large $%
\ell $ as $\ell \rightarrow \infty $ can be written as

\begin{equation}
f_{\nu }(\nu \omega ,\lambda )=J_{\nu }\left( \nu \omega \right) \;N_{\nu
}\left( \nu \omega \lambda \right) -J_{\nu }\left( \nu \omega \lambda
\right) \;N_{\nu }\left( \nu \omega \right)
\end{equation}%
Where $\lambda =\frac{a}{b}\ (b>a)$.

Then, the scalar Casimir energy between the closely spaced two concentric
spheres can be written as

\begin{eqnarray}
E_{C} &=&\sum_{\ell =0}^{\infty }\nu \sum_{n=1}^{\infty }\;\omega _{n\ell
}+\sum_{\ell =0}^{\infty }\nu \sum_{n=1}^{\infty }\sqrt{\left( \frac{n\pi }{%
b-a}\right) ^{2}+\frac{\nu ^{2}}{ab}}\ \   \notag \\
&&  \notag \\
&=&\overline{E}_{C}+\widetilde{E}_{C}\;.
\end{eqnarray}%
Where $\omega _{n\ell }$ is the root of the frequency equation given in Eq.
(12).

We consider the first sum defined in Eq. (13). This divergent expression can
be rendered finite by the use of a cutoff or convergence factor. Then$\;$we
define the first sum,

\begin{eqnarray}
\overline{E}_{C} &=&\sum_{\ell =0}^{\infty }\left( \ell +\frac{1}{2}\right)
\sum_{n=1}^{\infty }\;\omega _{n\ell }\;e^{-\alpha \omega _{n\ell }}  \notag
\\
&&  \notag \\
&=&\sum_{\ell =0}^{\infty }\left( \ell +\frac{1}{2}\right) \;S_{\ell }\;,
\end{eqnarray}%
where the factor of $e^{-\alpha \omega _{n\ell }}$ plays the role of an
exponential cutoff function, and $S_{\ell }=\sum_{n=1}^{\infty }\;\omega
_{n\ell }\;e^{-\alpha \omega _{n\ell }}$\ is generated by the frequency
equation (12). To evaluate the sum $S_{\ell }$, we use the integral
representation from the Cauchy's theorem \cite{Hag, Mozcan1,Mazzitelli} that
for two functions $f_{\ell }(z)$ and $\phi (z)$ analytic within a closed
contour C in which $f_{\ell }(z)$ has isolated zeros at $%
x_{1,}x_{2},x_{3,.........,}x_{n}\;,$

\begin{equation}
\frac{1}{2\pi i}\oint_{C}dz\;\phi (z)\;\frac{d}{dz}\ln f_{\ell
}(z)=\sum_{j}\phi (x_{j})\;.
\end{equation}
We choose $\phi (z)=z\;e^{-\alpha z}$ where $\alpha $ is a real positive
constant thus leads to

\begin{equation}
\frac{1}{2\pi i}\oint_{C}dz\;e^{-\alpha z}\;z\frac{d}{dz}\ln f_{\ell
}(z)=\sum_{j}z_{j}\;e^{-\alpha z_{j}}\;.
\end{equation}%
Using this result to replace the sum $S_{\ell }$ by a contour integral, the
first term of the Casimir energy becomes

\begin{equation}
\overline{E}_{C}=\sum_{\ell =0}^{\infty }\left( \ell +\frac{1}{2}\right) \;%
\frac{1}{2\pi i}\oint_{C}dz\;e^{-\alpha z}\;z\frac{d}{dz}\ln f_{\nu }(\nu
z)\;,
\end{equation}%
where the frequency function $f_{\nu }(\nu z)$ is given Eq. (12). The
contour $C$ encloses all the positive roots of the equation $f_{\nu }(\nu
z)=0$. This contour can be conveniently broken into three parts \cite%
{Hag,Mozcan1, Mazzitelli}. These consist of a circular segment $C_{\Gamma }$
and two straight line segments $\Gamma _{1}$ and $\Gamma _{2}$ forming an
angle $\phi $ and $\pi -\phi $ with respect to the imaginary axis. When the
radius $\Gamma $ is fixed, the contour $C_{\Gamma }$ encloses a finite
number of roots of the equation $f_{\nu }(\nu z)=0$. Since the sum of these
roots is obviously infinite, the radius $\Gamma $ is a regularization
parameter, taking the limit $\Gamma \rightarrow \infty $ ( when $\alpha >0\;$%
) means the removal of the regularization, the contribution of $C_{\Gamma }$
vanishes provided that $\phi \neq 0$. Hence the exponential cutoff function
in the Cauchy integral plays the role of the eliminate of the contribution
to the circular part of the contour integral. Taking the contributions along 
$\Gamma _{1}$ and $\Gamma _{2}$ which are complex conjugates of each other,
then Eq. (16) becomes \cite{Nest1, Hag}

\begin{equation}
\overline{E}_{C}=-\frac{1}{\pi }\lim_{\alpha \rightarrow 0}\sum_{\ell
=0}^{\infty }\nu \;\mathbf{\func{Re}}\;e^{-i\phi }\;\int_{0}^{\infty
}dy\;e^{-i\alpha ye^{-i\phi }}\;y\frac{d}{dy}\ln f_{\nu }(\nu ye^{-i\phi
};a,b)\;.\;
\end{equation}%
Where

\begin{equation}
f_{\nu }(\nu ye^{-i\phi };a,b)=-\frac{2}{\pi }\left[ I_{\nu }\left( \nu
ye^{-i\phi }b\right) K_{\nu }\left( \nu ye^{-i\phi }a\right) -I_{\nu }\left(
\nu ye^{-i\phi }a\right) K_{\nu }\left( \nu ye^{-i\phi }b\right) \right] \;
\end{equation}

Now we calculate the integral given in Eq. (18). Defining $\lambda =\frac{a}{%
b}\;(b>a)$ and after rescaling integral variable with $yb\rightarrow y$ in
equation $\overline{E}_{C}$, one obtains

\begin{equation}
\overline{E}_{C}=-\frac{1}{\pi b}\lim_{\alpha \rightarrow 0}\sum_{\ell
=0}^{\infty }\nu \;\mathbf{\func{Re}}\;e^{-i\phi }\;\int_{0}^{\infty
}dy\;e^{-i\alpha ye^{-i\phi }/b}\;y\frac{d}{dy}\ln f_{\nu }(\nu ye^{-i\phi
},\lambda )\;.
\end{equation}%
Where $f_{\nu }(\nu ye^{-i\phi },\lambda )=-\frac{2}{\pi }\left[ I_{\nu
}(\nu ye^{-i\phi })K_{\nu }(\nu ye^{-i\phi }\lambda )-I_{\nu }(\nu
ye^{-i\phi }\lambda )K_{\nu }(\nu ye^{-i\phi })\right] $ stems from Eq. (19).

Now we use the Lommel's expansions\ or the multiplication theorem for the
function of $f_{\nu }(\nu ye^{-i\phi },\lambda )$ \cite{Mozcan1,Watson}.
Thus, we have

$\;$%
\begin{eqnarray}
f_{\nu }(z,\lambda ) &=&I_{\nu }\left( z\right) \;K_{\nu }\left( z\lambda
\right) -I_{\nu }\left( z\lambda \right) \;K_{\nu }\left( z\right)  \notag \\
&&  \notag \\
&=&\lambda ^{-\nu }\sum_{k=0}^{\infty }\frac{\left( \lambda ^{2}-1\right)
^{k}}{k!2^{k}}z^{2k-\nu }  \notag \\
&&\times \left\{ I_{\nu }\left( z\right) \;\left( \frac{d}{z\;dz}\right)
^{k}\left\{ z^{\nu }K_{\nu }\left( z\right) \right\} -K_{\nu }\left(
z\right) \;\left( \frac{d}{z\;dz}\right) ^{k}\left\{ z^{\nu }I_{\nu }\left(
z\right) \right\} \right\} .  \notag \\
&&
\end{eqnarray}%
Where $\left\vert \lambda ^{2}-1\right\vert <1$. Applying the uniform with
respect to $z$ asymptotics for the modified Bessel functions at large $\nu $%
, after long calculations \cite{Mozcan1}, we obtain

\begin{eqnarray}
\nu y\;\frac{d}{dy}\ln \;f_{\nu }(\nu ye^{-i\phi },\lambda ) &=&\frac{\left(
1-\lambda ^{2}\right) ^{2}}{12}\left( \nu ye^{-i\phi }\right) ^{2}+\frac{%
\left( 1-\lambda ^{2}\right) ^{3}}{24}\left( \nu ye^{-i\phi }\right) ^{2} 
\notag \\
&&  \notag \\
&&+\frac{\left( 1-\lambda ^{2}\right) ^{4}}{720}\left[ \left( \nu ye^{-i\phi
}\right) ^{2}(-\nu ^{2}+19)-\left( \nu ye^{-i\phi }\right) ^{4}\right] 
\notag \\
&&  \notag \\
&&-\frac{\left( 1-\lambda ^{2}\right) ^{5}}{1440}\left[ 3\left( \nu
ye^{-i\phi }\right) ^{2}(\nu ^{2}-9)+2\left( \nu ye^{-i\phi }\right) ^{4}%
\right]  \notag \\
&&  \notag \\
&&+\frac{\left( 1-\lambda ^{2}\right) ^{6}}{120960}\left[ \left( \nu
ye^{-i\phi }\right) ^{2}(4\nu ^{4}-290\nu ^{2}+1726)+\left( \nu ye^{-i\phi
}\right) ^{4}(8\nu ^{2}-149)\right.  \notag \\
&&  \notag \\
&&+\left. 4\left( \nu ye^{-i\phi }\right) ^{6}\right] +\left[ \text{Terms in
even powers of }\left( \nu ye^{-i\phi }\right) \right] ,
\end{eqnarray}%
Inserting Eq. (22) into Eq. (20) and using the following integral result

\begin{eqnarray}
I(2n) &=&e^{-i\phi }\int_{0}^{\infty }dy\;e^{-i\alpha ye^{-i\phi
}/b}\;\left( \nu ye^{-i\phi }\right) ^{2n}  \notag \\
&&  \notag \\
&=&i\;\left( -\right) ^{n+1}\left( 2n\right) !\ \nu ^{2n}\;\left( \frac{b}{%
\alpha }\right) ^{2n+1}\;,\;\;\;\;\text{where }n=0,1,2,3...
\end{eqnarray}%
then Eq. (20) becomes

\begin{eqnarray}
\overline{E}_{C} &=&-\frac{1}{\pi b}\;\lim_{\alpha \rightarrow 0}\sum_{\ell
=0}^{\infty }\;\mathbf{\func{Re}\;}\left\{ i\;\frac{\left( 1-\lambda
^{2}\right) ^{2}}{6}\nu ^{2}\left( \frac{b}{\alpha }\right) ^{3}+i\;\frac{%
\left( 1-\lambda ^{2}\right) ^{3}}{12}\nu ^{2}\left( \frac{b}{\alpha }%
\right) ^{3}\right.  \notag \\
&&  \notag \\
&&+i\;\frac{\left( 1-\lambda ^{2}\right) ^{4}}{720}\left[ 2(-\nu ^{4}+19\nu
^{2})\left( \frac{b}{\alpha }\right) ^{3}+24\nu ^{4}\left( \frac{b}{\alpha }%
\right) ^{5}\right]  \notag \\
&&  \notag \\
&&+i\;\frac{\left( 1-\lambda ^{2}\right) ^{5}}{720}\left[ -6(\nu ^{4}-9\nu
^{2})\left( \frac{b}{\alpha }\right) ^{3}+48\nu ^{4}\left( \frac{b}{\alpha }%
\right) ^{5}\right]  \notag \\
&&  \notag \\
&&+i\;\frac{\left( 1-\lambda ^{2}\right) ^{6}}{120960}\left[ (8\nu
^{6}-580\nu ^{4}+3452\nu ^{2})\left( \frac{b}{\alpha }\right) ^{3}-24(8\nu
^{6}-149\nu ^{4})\left( \frac{b}{\alpha }\right) ^{5}+2880\nu ^{6}\left( 
\frac{b}{\alpha }\right) ^{7}\right]  \notag \\
&&  \notag \\
&&+\left. \left[ \text{Terms in imaginary number and even powers of }\nu %
\right] \right\} \;.
\end{eqnarray}%
All terms in the above equation have the singular term in the regulator
parameter $\alpha $ and purely imaginary. Taking the real part of the
parenthesis, thus it leaves the zero result.

\begin{equation}
\overline{E}_{C}=0\;.
\end{equation}

The meaning of this result is that there is no contribution from $\ell
\rightarrow \infty $ modes for the Casimir energy between two concentric
spheres \cite{Mozcan1}. Now we return to the second sum given in Eq. (13),
included high eigenfrequency modes i.e. $\omega \rightarrow \infty $ at
fixed $\ell $.

\begin{equation}
\widetilde{E}_{C}\;=\sum_{\ell =0}^{\infty }\nu \sum_{n=1}^{\infty }\sqrt{%
\left( \frac{n\pi }{b-a}\right) ^{2}+\frac{\nu ^{2}}{ab}}.
\end{equation}%
Where $\nu =\left( \ell +\frac{1}{2}\right) .$ This divergent sum can be
regularized by using the Abel-Plana sum formula. Before we proceed, we
recall the Abel-Plana formula gives an expression for the difference between
the sum and corresponding integral, which could be given as \cite{Mosbook,
Mos1}

\begin{eqnarray}
\text{Reg}\left[ \sum_{n=1}^{\infty }f(n)\right] &=&  \notag \\
&&  \notag \\
\dsum\limits_{n=0}^{\infty }f\left( n\right) -\dint\limits_{0}^{\infty
}f(x)dx &=&\frac{1}{2}f(0)+i\dint\limits_{0}^{\infty }\frac{f(it)-f(-it)}{%
e^{2\pi t}-1}dt.
\end{eqnarray}%
Where $f\left( z\right) $ is an analytic function in the right half plane
and Reg refers to the regularized value of the sum. The other useful
Abel-Plana sum formula for the half integer number is

\begin{equation}
\dsum\limits_{n=0}^{\infty }f\left( n+\frac{1}{2}\right)
=\dint\limits_{0}^{\infty }f(x)dx-i\dint\limits_{0}^{\infty }\frac{%
f(it)-f(-it)}{e^{2\pi t}+1}dt.
\end{equation}

We can rewrite the sum given in Eq. (26) using the Abel-Plana sum formula,
which leads to

\begin{eqnarray}
\widetilde{E}_{C}\; &=&\sum_{\ell =0}^{\infty }\nu \ \ \text{Reg}\left[
\sum_{n=1}^{\infty }\sqrt{\left( \frac{n\pi }{b-a}\right) ^{2}+\frac{\nu ^{2}%
}{ab}}\right]  \notag \\
&&  \notag \\
&=&-\frac{1}{2\sqrt{ab}}\sum_{\ell =0}^{\infty }\ \nu ^{2}-2\sum_{\ell
=0}^{\infty }\nu \ \dint\limits_{\frac{\nu \xi }{2\pi }}^{\infty }\left[
\left( \frac{t\pi }{b-a}\right) ^{2}-\frac{\nu ^{2}}{ab}\right] ^{\frac{1}{2}%
}\frac{dt}{e^{2\pi t}-1}\ .
\end{eqnarray}%
where $\xi =\frac{2d}{\sqrt{ab}}$ and $d=b-a$. The first divergent sum in
Eq. (29) can be removed by using the Hurwitz zeta function \cite{Elizalde
book}

\begin{eqnarray}
\sum_{\ell =0}^{\infty }\left( \ell +\frac{1}{2}\right) ^{2} &=&\zeta \left(
-2,\frac{1}{2}\right)  \notag \\
&=&0.
\end{eqnarray}%
Thus Eq. (29) becomes

\begin{eqnarray}
\widetilde{E}_{C} &=&-2\sum_{\ell =0}^{\infty }\ F\left( \nu \right)  \notag
\\
\text{where }F\left( \nu \right) &=&\nu \ \dint\limits_{\frac{\nu \xi }{2\pi 
}}^{\infty }\left[ \left( \frac{t\pi }{b-a}\right) ^{2}-\frac{\nu ^{2}}{ab}%
\right] ^{\frac{1}{2}}\frac{dt}{e^{2\pi t}-1}.
\end{eqnarray}%
Using the half integer Abel-Plana sum formula given in Eq. (28), one obtains

\begin{equation}
\widetilde{E}_{C}=-\frac{1}{4\pi }\frac{\left( \sqrt{ab}\right) ^{2}}{d^{3}}%
\zeta \left( 4\right) +2i\dint\limits_{0}^{\infty }\frac{F(it)-F(-it)}{%
e^{2\pi t}+1}dt.
\end{equation}%
Where $\zeta \left( s\right) $ is the Riemann zeta function. The divergent
term occurs in the second term of the Eq. (32). We avoid this problem by
substituting $\nu -i\in $ for $\nu $ with the understanding that we will let 
$\in \rightarrow 0$ at the end. Thus the Casimir energy for a massless
scalar field between the closely spaced two concentric spheres is obtained

\begin{eqnarray}
E_{C} &=&\overline{E}_{C}+\widetilde{E}_{C}\;  \notag \\
&=&-\frac{1}{4\pi }\frac{\left( \sqrt{ab}\right) ^{2}}{d^{3}}\zeta \left(
4\right) \left[ 1+\frac{1}{12}\eta ^{2}\frac{\zeta \left( 2\right) }{\zeta
\left( 4\right) }\right] ,  \notag \\
&&  \notag \\
E_{C} &=&-\frac{\pi ^{3}}{360}\frac{ab}{d^{3}}\left[ 1+\frac{5}{4\pi ^{2}}%
\frac{d^{2}}{ab\ }\right] .  \notag \\
&&
\end{eqnarray}%
Where $\eta =\frac{d}{\sqrt{ab}}$. And known formula $\zeta \left( 4\right) =%
\frac{\pi ^{4}}{90}$ and $\zeta \left( 2\right) =\frac{\pi ^{2}}{6}$. The
main contribution to the Casimir energy is given by the second term which
represents the high frequency modes at fixed $\ell $ between the closely
spaced two concentric spheres for a massless scalar field. The Casimir
energy per unit surface area on the inner sphere (the total surface area $%
A=4\pi a^{2})$ can be written as

\begin{equation}
\frac{E_{C}}{A}=-\frac{1}{16\pi ^{2}}\left( \frac{\sqrt{ab}}{a}\right) ^{2}%
\frac{\zeta \left( 4\right) }{d^{3}}\left[ 1+\frac{1}{12}\eta ^{2}\frac{%
\zeta \left( 2\right) }{\zeta \left( 4\right) }\right]
\end{equation}%
This result is interest at the limiting case which is narrow slit is defined
by $\eta =\frac{d}{\sqrt{ab}}\ll 1$ \cite{Brevik1}. We easily analysis that
in the limit $b\rightarrow a$ ($i.e.\eta \rightarrow 0$ and $\frac{\sqrt{ab}%
}{a}\rightarrow 1$) which means that the close separation between two
spheres one finds that the leading term of the Casimir energy per unit area
can be written as

\begin{equation}
\frac{E_{C}}{A}=-\frac{1}{16\pi ^{2}}\frac{\zeta \left( 4\right) }{d^{3}}.
\end{equation}%
This result is exactly the same as the Casimir energy of the parallel plates
for a massless scalar field \cite{JanAmbjorn}. Thus our approach developed
here has been the satisfactory check.

\section{CASIMIR ENERGY BETWEEN TWO CONCENTRIC HALF SPHERES}

The our approach can be easily employed to the computation of the Casimir
energy of a massless scalar field between the closely spaced two concentric
half spheres. In this case of our spherical boundary \cite{Mozcan1} the sum
in Eq. (13) can be written as

\begin{equation}
E_{C}=\frac{1}{2}\sum_{\ell =1}^{\infty }\ell \sum_{n=1}^{\infty }\;\omega
_{n\ell }+\frac{1}{2}\sum_{\ell =1}^{\infty }\ell \sum_{n=1}^{\infty }\sqrt{%
\left( \frac{n\pi }{b-a}\right) ^{2}+\frac{\nu ^{2}}{ab}}\ .
\end{equation}%
Where $\nu =\ell +\frac{1}{2}$ and $\omega _{n\ell }$ is the root of the
transcendental equation given in Eq. (6). From Eq. (15) we obtain

\begin{eqnarray}
E_{C} &=&\overline{E}_{C}+\widetilde{E}_{C}\;  \notag \\
&&  \notag \\
&=&-\frac{1}{2\pi b}\lim_{\alpha \rightarrow 0}\sum_{\ell =1}^{\infty
}\left( \nu -\frac{1}{2}\right) \;\mathbf{\func{Re}}\;e^{-i\phi
}\;\int_{0}^{\infty }dy\;e^{-i\alpha ye^{-i\phi }/b}\;y\frac{d}{dy}\ln
f_{\nu }(\nu ye^{-i\phi },\lambda )  \notag \\
&&+\frac{1}{2}\sum_{\ell =1}^{\infty }\left( \nu -\frac{1}{2}\right)
\sum_{n=1}^{\infty }\sqrt{\left( \frac{n\pi }{b-a}\right) ^{2}+\frac{\nu ^{2}%
}{ab}}
\end{eqnarray}%
Where $y\frac{d}{dy}\ln f_{\nu }(\nu ye^{-i\phi },\lambda )$ given in eq.
(22). The first term $\left( =\overline{E}_{C}\right) $ is equal to zero
from the Eq. (25). Thus the Casimir energy between the closely spaced two
concentric half spheres becomes

\begin{equation}
E_{C}=\;\frac{1}{2}\sum_{\ell =1}^{\infty }\left( \nu -\frac{1}{2}\right)
\sum_{n=1}^{\infty }\sqrt{\left( \frac{n\pi }{b-a}\right) ^{2}+\frac{\nu ^{2}%
}{ab}}
\end{equation}

Using the Abel-plana sum formula one obtains

\begin{eqnarray}
E_{C} &=&-\;\frac{1}{4\sqrt{ab}}\sum_{\ell =1}^{\infty }\left( \nu ^{2}-%
\frac{\nu }{2}\right) -\frac{d}{\pi ab}\sum_{\ell =1}^{\infty }F\left( \nu
\right) , \\
\text{where }F\left( \nu \right) &=&\left( \nu ^{3}-\frac{\nu ^{2}}{2}%
\right) \ \dint\limits_{1}^{\infty }\left( y^{2}-1\right) ^{1/2}\frac{dy}{%
e^{2\pi y}-1}\text{ , \ \ \ and }d=b-a.\text{\ \ \ }  \notag
\end{eqnarray}%
Again using the half integer Abel-Plana sum formula for the second sum in
Eq. (39), we obtain

\begin{equation}
E_{C}=-\frac{1}{8\pi }\frac{\left( \sqrt{ab}\right) ^{2}}{d^{3}}\zeta \left(
4\right) \left[ 1-\frac{\pi }{4}\eta \frac{\zeta \left( 3\right) }{\zeta
\left( 4\right) }+\frac{\pi }{24}\eta ^{3}\frac{1}{\zeta \left( 4\right) }+%
\frac{1}{12}\eta ^{2}\frac{\zeta \left( 2\right) }{\zeta \left( 4\right) }-%
\frac{1}{4\pi }\eta ^{3}\frac{1}{\zeta \left( 4\right) }\right] .
\end{equation}%
We have used the Hurwitz zeta function i.e. $\zeta \left( -m,q\right) =-%
\frac{B_{m+1}\left( q\right) }{m+1},$ \ $m=0,1,2,3,...$ where $B_{m+1}\left(
q\right) $ is the Bernoulli polynomials.

The Casimir energy of a massless scalar field between the closely spaced two
concentric half spheres per unit half surface area on the inner sphere (the
half surface area $A=2\pi a^{2})$ can be written as

\begin{equation}
\frac{E_{C}}{A}=-\frac{1}{16\pi }\left( \frac{\sqrt{ab}}{a}\right) ^{2}\frac{%
\zeta \left( 4\right) }{d^{3}}\left[ 1-\frac{\pi }{4}\eta \frac{\zeta \left(
3\right) }{\zeta \left( 4\right) }+\frac{\pi }{24}\eta ^{3}\frac{1}{\zeta
\left( 4\right) }+\frac{1}{12}\eta ^{2}\frac{\zeta \left( 2\right) }{\zeta
\left( 4\right) }-\frac{1}{4\pi }\eta ^{3}\frac{1}{\zeta \left( 4\right) }%
\right] .
\end{equation}%
Where $\zeta \left( s\right) \ $is the Riemann zeta function.

Taking the small separation limit $\left( b\rightarrow a\text{ and }\eta
\rightarrow 0\right) $ one obtains

\begin{equation}
\frac{E_{C}}{A}=-\frac{1}{16\pi }\frac{\zeta \left( 4\right) }{d^{3}}.
\end{equation}%
This result coincidences with the Casimir energy of the parallel plates for
the massless scalar field \cite{JanAmbjorn}.

\section{\protect\bigskip CONCLUSION}

\qquad In the present paper we considered the quantum vacuum energy for a
massless scalar field between the closely spaced two concentric spheres, and
between two concentric half spheres at small separations. The annular region
in our geometry was considered since all massless scalar fields in the
region $r<a$ and $r>b$ are equal to zero. The Casimir energy for a massless
scalar field in annular region was evaluated by a direct mode summation
method. We have used the explicit expression for the frequency equations
which include the product of Bessel functions. In order to evaluate the
product of Bessel function expansion for large order in our approach, the
Lommel's expansions has been used in the Casimir problem for the first time.
These calculation is a direct application of the principle of argument from
the complex integral with the cutoff function, in direct analogy to the
Casimir calculation, together with a modification of the contour in order to
ensure the convergence of the Cauchy integral expressions. Moreover, the
product of Bessel function expansion for the large argument is defined by
evenly spaced eigenfrequency spectrum. Although we use the Abel-Plana sum
formula, divergent terms appear in our calculations. To remove this
divergence we have applied the formal technique of the Zeta function
regularization. Thus, our calculation implies that further regularization is
not needed in order for the finite values for the vacuum energy to be
obtained for given boundary conditions.

We are particularly interested in calculating the Casimir energy between
spherical surfaces close to each other is that any approximation technique
is not needed. The interesting point of our calculations is that all
contributions in the Casimir energy for a massless scalar field comes from
the higher frequencies for \ fixed $\ell $ between two surfaces boundary
conditions. $\ell \rightarrow \infty $ frequency modes contribution in the
Casimir energy is zero for close separation of annular region in present
geometries.

Although the Casimir energy sign for the massless scalar field on spherical
shell with the Dirichlet boundary condition is positive, our analysis
reveals the negative sign of the Casimir energy between the closely spaced
two concentric spheres, and between two concentric half spheres at small
separations. Then, the Casimir force for a massless scalar field between two
concentric spheres, and between two concentric half spheres are always
attractive which is the same per unit area on a pair of parallel plates for
a massless scalar field \cite{JanAmbjorn}. Moreover, the leading term of the
Casimir energy between the closely spaced two concentric spheres given in
Eq. (33) agrees with that obtained the Casimir effect between two spheres at
small separation by using the functional determinant or multiple scattering
approach\cite{zaheer,teo2}.

Closing, it is worth noting that, as far as we know, such boundary
conditions with between the closely spaced two concentric spheres has been
considered in the Casimir problem with the use mode sum technique for the
first time. As far as we know this result is obtained here for the first
time.

For future works, it would be interesting to consider the electrodynamics
Casimir energy between the closely spaced two concentric spheres with those
obtained.


\begin{thebibliography}{99}
\bibitem{Cas} H. B. G. Casimir, Proc. K. Ned. Acad. Wet. \textbf{51, }793
(1948)\textbf{.}

\bibitem{Spaar} M. J. Sparnaay, Physica \textbf{24}, 751 (1958).

\bibitem{Lam} S. K. Lamoreaux, Phys. Rev. Lett. \textbf{78}, 5 (1997).

\bibitem{mohideen} G. L. Klimchitskaya, U. Mohideen and V. M. Mostepanenko,
Rev. Mod. Phys. \textbf{81}, 1827 (2009).

\bibitem{Mozcan2} Mustafa \"{O}zcan, Class. Quantum Grav. \textbf{23, }5531
(2006).

\bibitem{Mozcan3} S. S. Bayin and M. \"{O}zcan, Phys. Rev. \textbf{D 48},
2806 (1993).

\bibitem{Miltonbook} K. A. Milton, Physical Manifestations of Zero-Point
Energy The Casimir Effect (World Scientific 2001).

\bibitem{Gunter} G\"{u}nter Plunien, Berndt M\"{u}ller and Walter Greiner,
Physics Reports \textbf{134}, 87 (1986).

\bibitem{Mosbook} V. M. Mostepanenko and N. N. Trunov, \textit{The Casimir
Effect and its Applications} (Oxford University Press, New York, 1997).

\bibitem{Mos1} M. Bordag, U. Mohideen and V. M. Mostepanenko, Phys. Rep. 
\textbf{353}, 1 (2001).

\bibitem{Boyer} T. H. Boyer, Phys. Rev. \textbf{174, }1764 (1968).

\bibitem{Milton1} K. A. Milton, L. L. DeRaad, Jr. , and J. Schwinger, Ann.
Phys. (NY) \textbf{115}, 388 (1978).

\bibitem{Nest1} V. V. Nesterenko and L. G. Pirozhenko, Phys. Rev. D \textbf{%
57}, 1284 (1998).

\bibitem{Hag} M. E. Bowers and C. R. Hagen, Phys. Rev. D \textbf{59}, 025007
(1999).

\bibitem{bordag} M. Bordag, Phys. Rev. D \textbf{73}, 125018 (2006).

\bibitem{emig1} T. Emig, N. Graham, R. L. Jaffe, and M. Kardar, Phys. Rev.
Lett. \textbf{99}, 170403 (2007).

\bibitem{emig2} T. Emig, N. Graham, R. L. Jaffe, and M. Kardar, Phys. Rev. D 
\textbf{77}, 025005 (2008).

\bibitem{emig3} T. Emig and R. L. Jaffe, J. Phys. A: Math. Theor. \textbf{41}%
, 164001 (2008).

\bibitem{kenneth} O. Kenneth and I. Klich, Phys. Rev. B \textbf{78}, 014103
(2008).

\bibitem{zaheer} Saad Zaheer, S. J. Rahi, T. Emig, and R. L. Jaffe, Phys.
Rev. A \textbf{81}, 030502 (2010); Phys. Rev. A \textbf{82}, 052507 (2010).

\bibitem{rahi} S. J. Rahi, T. Emig, N. Graham, R. L. Jaffe, and M. Kardar,
Phys. Rev. D \textbf{80}, 085021 (2009).

\bibitem{Brevik1} J. S. Hoye, I. Brevik, and J. B. Aarseth, Phys Rev. E 
\textbf{63}, 051101 (2001); I. Brevik, E. K. Dahl, and G. O. Myhr, J. Phys.
A: Math. Gen. \textbf{38}, L49 (2005).

\bibitem{Miltao} M. S. R. Miltao, Phys. Rev. D \textbf{78}, 065023 (2008).

\bibitem{teo1} L. P. Teo, Phys. Rev. D \textbf{82}, 085009 (2010).

\bibitem{IHD} H. Ahmedov and I. H. Duru, J. of Math. Phys. \textbf{12}, 5487
(2003).

\bibitem{Mozcan1} Mustafa \"{O}zcan, Physics Letters A 344, 307 (2005).

\bibitem{Jiang} W. Z. Jiang, Z. X. Wang, D. J. Fu, H. B. Ai and Z. Y. Zhu,
Physics Letters A \textbf{315}, 273 (2003).

\bibitem{Tatur} K. Tatur and L. M. Woods, Physics Letters A \textbf{372},
6705 (2008).

\bibitem{Mazzitelli} F. D. Mazzitelli, M. J. Sanchez, N. N. Scoccola and J.
von Stecher, Phys. Rev. A 67, 013807 (2003).

\bibitem{Funaro} Daniele Funaro, arXiv: physics.gen-ph/0906.1874v1.

\bibitem{Abramowitz} M. Abramowitz and I. A. Stegun, \textit{Hanbook of
Mathematical Functions} (National Bureau of Standards, Washington, D. C.,
1964).

\bibitem{Watson} G. N. Watson, \textit{A treatise on the THEORY OF BESSEL
FUNCTIONS} (Cambridge Press, Cambridge, England, 1966).

\bibitem{Elizalde book} E. Elizalde, S. D. Odintsov, A. Romeo, A. A.
Bytsenko, and S. Zerbini, \textit{Zeta Regularization Techniques with
applications (World scientific, 1994).}

\bibitem{JanAmbjorn} Jam Ambjorn and Stephen Wolfram, Annals of Physics 147,
1 (1983).

\bibitem{teo2} L. P. Teo, Phys Rev. \textbf{D} 85, 045027 (2012); A. Bulgac,
P. Magierski and A. Wirzbu, Phys Rev. \textbf{D} 73, 0250007 (2006).
\end{thebibliography}
\end{document}